\documentclass[english]{extarticle}
\usepackage[T1]{fontenc}
\usepackage[latin9]{inputenc}
\usepackage{amsbsy}
\usepackage{amssymb}
\usepackage{esint}

\makeatletter
\usepackage{fullpage}

\makeatother

\usepackage{babel}
\begin{document}
\title{Abelian symmetry and the Palatini variation}
\author{James T. Wheeler\thanks{Utah State University, Logan, UT 84322, jim.wheeler@usu.edu}}
\maketitle
\begin{abstract}
Independent variation of the metric and connection in the Einstein-Hilbert
action, called the Palatini variation, is generally taken to be equivalent
to the usual formulation of general relativity in which only the metric
is varied. However, when an abelian symmetry is allowed for the connection,
the Palatini variation leads to an integrable Weyl geometry, not Riemannian.
We derive this result using two possible metric/connection pairs:
(1) the metric and general coordinate connection and (2) the solder
form and local Lorentz spin connection of Poincarè gauge theory. Both
lead to the same conclusion. Finally, we relate our work to other
treatments in the literature.
\end{abstract}

\section{The Palatini variation}

General relativity describes spacetimes, $\left(\mathcal{M},g\right)$,
where $\mathcal{M}$ is a Riemannian manifold and $g$ is a Lorentzian
metric. The field equation follows by metric variation of the action
functional
\[
S_{GR}\left[g\right]=\int R\sqrt{-g}d^{4}x
\]
where $R$ is the scalar curvature computed from the metric compatible
Christoffel connection. Sources are included by adding the action
for any generally coordinate invariant matter action to $S_{GR}$,
\[
S=S_{GR}+S_{Matter}
\]

The beginning of an alternative variation dates back to a 1919 paper
by \cite{Palatini} and was brought to its current formulation by
Einstein \cite{Einstein} (see \cite{Frncaviglia} for the interesting
history leading Einstein to the connection variation, and Appendix
I for Einstein's calculation). The alternative formulation showed
that the assumption of the metric compatible connection could be replaced
by varying the metric and connection independently, using the action,
\begin{equation}
S_{P}\left[g,\hat{\Gamma}\right]=\int\hat{R}\sqrt{-g}d^{4}x\label{Palatini action}
\end{equation}
Here $\hat{\Gamma}$ is any symmetric connection, $\hat{\Gamma}_{\;\;\;\mu\nu}^{\alpha}=\hat{\Gamma}_{\;\;\;\nu\mu}^{\alpha}$.
This symmetry condition is preserved by changes of coordinates because
the inhomogeneous term from a general coordinate transformation is
symmetric.

Notice that in treating the connection independently, we consider
spacetime to be a triple, $\left(\mathcal{M},g,\hat{\Gamma}\right)$.

Although the independent variable $\hat{\Gamma}_{\;\;\;\mu\nu}^{\alpha}$
is assumed to be a general symmetric connection, this is not the form
taken by the connection for an abelian symmetry, which carries a weight
factor and can apply nontrivially to scalars as well as vectors. For
the connection variation to be complete, a more general expression
is required. In the remainder of this Section, we carry out the usual
Palatini variation of $S_{P}$, then show the altered effect of an
abelian covariance to the derivation. We find that including an abelian
term in the connection results in an integrable Weyl geometry.

In Section 2 we study the Palatini variation in Poincarè gauge theory,
where the independent variables are the solder form and the spin connection
instead of the metric $g_{\mu\nu}$ and $\hat{\Gamma}_{\;\;\;\mu\nu}^{\alpha}$.
The details differ in interesting ways but the end result is the same--when
the possibility of an abelian symmetry is included in the variation
of the spin connection, we obtain an integrable Weyl geometry.

As in our investigation, Einstein's original development of the Palatini
variation leads to the introduction of an additional vector field.
In the final Section we discuss the relationship between this vector
field and our inclusion of abelian symmetry. We conclude by noting
the differences between ours and some standard treatments of the Palatini
variation.

Throughout these notes, in order to distinguish coordinate and orthonormal
frames, Greek indices refer to any coordinate basis, and Latin indices
to any orthonormal basis. We do not use coordinate-free tensor notation
since this would unnecessarily complicate the notation.

\subsection{The standard Palatini variation\label{subsec:The-standard-Palatini}}

With the Palatini variation of Eq.(\ref{Palatini action}), the metric
variation becomes much simpler. Writing the metric dependence explicitly
and varying
\begin{eqnarray*}
\delta_{g}S_{P}\left[g,\hat{\Gamma}\right] & = & \delta_{g}\int\hat{R}_{\alpha\beta}g^{\alpha\beta}\sqrt{-g}d^{4}x\\
 & = & \delta_{g}\int\left(\hat{R}_{\alpha\beta}-\frac{1}{2}g_{\alpha\beta}\hat{R}\right)\delta g^{\alpha\beta}\sqrt{-g}d^{4}x
\end{eqnarray*}
gives the Einstein tensor,
\begin{eqnarray}
\hat{R}_{\alpha\beta}-\frac{1}{2}g_{\alpha\beta}\hat{R} & = & 0\label{Metric variation}
\end{eqnarray}
Here $\hat{R}_{\alpha\beta}$ is the Ricci tensor computed from $\hat{\Gamma}$,
but it is only after varying $\hat{\Gamma}$ that we know what connection
to use.

The the connection variation gives
\begin{eqnarray}
\delta_{\Gamma}S_{P}\left[g,\hat{\Gamma}\right] & = & \delta_{\Gamma}\int\hat{R}_{\alpha\beta}g^{\alpha\beta}\sqrt{-g}d^{4}x\nonumber \\
 & = & \int\left(\hat{D}_{\mu}\left(\delta\hat{\Gamma}_{\;\;\;\alpha\beta}^{\mu}\right)-\hat{D}_{\beta}\left(\delta\hat{\Gamma}_{\;\;\;\alpha\mu}^{\mu}\right)\right)g^{\alpha\beta}\sqrt{-g}d^{4}x\label{Initial connection variation}
\end{eqnarray}
At this point it is useful to write the covariant derivative as the
sum of a metric compatible piece and an additional, non-compatible
tensor.
\[
\hat{D}_{\mu}v^{\alpha}=\partial_{\mu}v^{\alpha}+v^{\beta}\hat{\Gamma}_{\;\;\;\beta\mu}^{\alpha}=\triangledown_{\mu}v^{\alpha}+v^{\beta}C_{\;\;\;\beta\mu}^{\alpha}
\]
so the connection variation becomes variation of the non-metric piece,
$\delta\hat{\Gamma}_{\;\;\;\alpha\beta}^{\mu}=\delta C_{\;\;\;\alpha\beta}^{\mu}$.
Here $\triangledown_{\mu}g_{\alpha\beta}=0$ by definition, implying
the usual Christoffel/Levi-Civita connection for $\nabla_{\mu}$.
The remaining tensor $C_{\;\;\;\beta\mu}^{\alpha}$ is intended to
characterize any further properties of the connection. We carry this
out in considerable detail for subsequent reference.

For the variation we need
\[
\hat{D}_{\mu}\left(\delta C_{\;\;\;\alpha\beta}^{\nu}\right)=\bigtriangledown_{\mu}\left(\delta C_{\;\;\;\alpha\beta}^{\nu}\right)+\left(\delta C_{\;\;\;\alpha\beta}^{\rho}\right)C_{\;\;\;\rho\mu}^{\nu}-\left(\delta C_{\;\;\;\rho\beta}^{\nu}\right)C_{\;\;\;\alpha\mu}^{\rho}-\left(\delta C_{\;\;\;\alpha\rho}^{\nu}\right)C_{\;\;\;\beta\mu}^{\rho}
\]
Taking the required contractions, the variation becomes
\begin{eqnarray*}
\delta_{\Gamma}S_{P}\left[g,\hat{\Gamma}\right] & = & \int\left(\triangledown_{\mu}\left(\delta C_{\;\;\;\alpha\beta}^{\mu}\right)-\bigtriangledown_{\beta}\left(\delta C_{\;\;\;\alpha\mu}^{\mu}\right)\right)g^{\alpha\beta}\sqrt{-g}d^{4}x\\
 &  & +\int\left(\delta C_{\;\;\;\alpha\beta}^{\rho}C_{\;\;\;\rho\mu}^{\mu}-\delta C_{\;\;\;\rho\beta}^{\mu}C_{\;\;\;\alpha\mu}^{\rho}-\delta C_{\;\;\;\alpha\rho}^{\mu}C_{\;\;\;\beta\mu}^{\rho}\right)g^{\alpha\beta}\sqrt{-g}d^{4}x\\
 &  & +\int\delta C_{\;\;\;\rho\nu}^{\nu}C_{\;\;\;\alpha\beta}^{\rho}g^{\alpha\beta}\sqrt{-g}d^{4}x
\end{eqnarray*}
The compatible part of each derivative in the variation is integrated
by parts and vanishes by metric-compatibility
\begin{eqnarray*}
\int\triangledown_{\mu}\left(\delta C_{\;\;\;\alpha\beta}^{\mu}\right)g^{\alpha\beta}\sqrt{-g}d^{4}x & = & -\int\delta C_{\;\;\;\alpha\beta}^{\mu}\triangledown_{\mu}\left(g^{\alpha\beta}\sqrt{-g}\right)d^{4}x=0\\
-\int\bigtriangledown_{\beta}\left(\delta C_{\;\;\;\alpha\mu}^{\mu}\right)g^{\alpha\beta}\sqrt{-g}d^{4}x & = & \int\delta C_{\;\;\;\alpha\mu}^{\mu}\bigtriangledown_{\beta}\left(g^{\alpha\beta}\sqrt{-g}\right)d^{4}x=0
\end{eqnarray*}
Collecting the remaining terms and setting the whole to zero,
\begin{eqnarray*}
0 & = & \int\delta C_{\;\;\;\sigma\lambda}^{\rho}\left(\delta_{\alpha}^{\sigma}\delta_{\beta}^{\lambda}C_{\;\;\;\rho\mu}^{\mu}-\delta_{\beta}^{\lambda}C_{\;\;\;\alpha\rho}^{\sigma}-\delta_{\alpha}^{\sigma}C_{\;\;\;\beta\rho}^{\lambda}+\delta_{\rho}^{\lambda}C_{\;\;\;\alpha\beta}^{\sigma}\right)g^{\alpha\beta}\sqrt{-g}d^{4}x
\end{eqnarray*}
from which we conclude
\begin{eqnarray*}
g^{\alpha\beta}\left(\delta_{\alpha}^{\sigma}\delta_{\beta}^{\lambda}C_{\;\;\;\rho\mu}^{\mu}-\delta_{\beta}^{\lambda}C_{\;\;\;\alpha\rho}^{\sigma}-\delta_{\alpha}^{\sigma}C_{\;\;\;\beta\rho}^{\lambda}+\delta_{\rho}^{\lambda}C_{\;\;\;\alpha\beta}^{\sigma}\right) & = & 0
\end{eqnarray*}
Carrying out the contractions
\begin{eqnarray*}
g^{\sigma\lambda}C_{\;\;\;\rho\mu}^{\mu}-C_{\quad\rho}^{\sigma\lambda}-C_{\quad\rho}^{\lambda\sigma}+\delta_{\rho}^{\lambda}C_{\quad\beta}^{\sigma\beta} & = & 0
\end{eqnarray*}

This is easily solved. Recalling the symmetry of the connection $C_{\;\;\;\alpha\beta}^{\mu}=C_{\;\;\;\beta\alpha}^{\mu}$,
there are only two independent contractions. From the $\sigma\lambda$
and $\lambda\rho$ contractions, the two traces must satisfy both
\begin{eqnarray*}
2C_{\;\;\;\sigma\rho}^{\sigma}+C_{\rho\quad\;\;\beta}^{\;\;\;\beta} & = & 0\\
3C_{\quad\beta}^{\sigma\beta} & = & 0
\end{eqnarray*}
and therefore both vanish. Substituting into the full field equation
and lowering indices we have
\begin{eqnarray*}
C_{\sigma\lambda\rho}+C_{\lambda\sigma\rho} & = & 0
\end{eqnarray*}
Finally, this succumbs to the usual technique of cycling the indices,
then adding the first two permutations and subtracting the third.
The result is the vanishing of the non-metric part of the connection
and we appear to have established metric compatibility.

\subsection{The connection of an abelian symmetry}

In addition to assuming symmetry of the connection, there is a further
hidden assumption. We noted above that $C_{\;\;\;\alpha\beta}^{\mu}$
is intended to account for all characteristics beyond metric compatibility,
but it fails to include the possibility of an abelian symmetry.

By contracting a connection such as $\hat{\Gamma}_{\;\;\;\alpha\beta}^{\mu}$
above with a small displacement $dx^{\beta}$ we see that
\[
M_{\;\;\;\alpha}^{\mu}=\hat{\Gamma}_{\;\;\;\alpha\beta}^{\mu}dx^{\beta}
\]
has the form of a linear transformation. The transformation characterizes
the relationship between components of a tangent vector in tangent
spaces separated by $dx^{\beta}$. By correcting for this change of
tangent basis in moving about a manifold, the covariant derivative
is able to separate the change in a physical vector field from the
arbitrariness of the coordinates. Parallel transport around a closed
loop therefore gives intrinsic geometric information--the curvature.

A linear transformation such as $M_{\;\;\;\alpha}^{\mu}$ is appropriate
for any non-abelian group of transformations. Linear representations
of non-abelian groups act on real or complex vector spaces, and must
be characterized by matrix transformations of dimension $n\geq2$.
To accomplish the Leibnitz rule for products of fields, we include
$k$ linear such transformations on tensors of rank $k$. For example,
the covariant derivative of the rank-2 metric is
\begin{eqnarray}
\hat{D}_{\mu}g_{\alpha\beta} & = & \partial_{\mu}g_{\alpha\beta}-g_{\rho\beta}\hat{\Gamma}_{\;\;\;\alpha\mu}^{\rho}-g_{\alpha\rho}\hat{\Gamma}_{\;\;\;\beta\mu}^{\rho}\label{Non-abelian derivative of the metric}
\end{eqnarray}

For an abelian group, the transformation is simple multiplication,
so that even scalars may provide nontrivial linear representations.
As a result, the connection for an abelian tranformation takes a different
form, acting nontrivially on \emph{weighted scalars}.

For example, a complex wave function under a $U\left(1\right)$ transformation
will transform as $\psi\rightarrow e^{i\alpha}\psi$, so the connection
required to make the $U\left(1\right)$ symmetry local acts on $\psi$
as
\[
\hat{D}_{\mu}\psi=\partial_{\mu}\psi-iA_{\mu}\psi
\]
and it follows that for a field transforming as $\chi_{\left(k\right)}\rightarrow\left(e^{i\alpha}\right)^{k}\chi_{\left(k\right)}$
(e.g., $\left(\psi\right)^{k}$) the derivative must include a \emph{weight}
$k$
\[
\hat{D}_{\mu}\chi_{\left(k\right)}=\partial_{\mu}\chi_{\left(k\right)}-ikA_{\mu}\chi_{\left(k\right)}
\]
The derivative $\hat{D}_{\mu}$ of $\chi_{\left(k\right)}$ is then
covariant under the combined transformation 
\begin{eqnarray*}
\chi_{\left(k\right)} & \rightarrow & e^{ik\varphi}\chi_{\left(k\right)}\\
A_{\alpha} & \rightarrow & A_{\alpha}+\partial_{\alpha}\varphi
\end{eqnarray*}

Derivations are transformations which are both linear and Leibnitz.
The weight is necessary in order to satisfy the Leibnitz rule. Thus,
for fields $\chi_{\left(k\right)}$ and $\psi_{\left(m\right)}$ of
weights $k$ and $m$ the weights are additive,
\begin{eqnarray*}
D_{\alpha}\left(\chi_{\left(k\right)}\psi_{\left(m\right)}\right) & = & \partial_{\alpha}\left(\chi_{\left(k\right)}\psi_{\left(m\right)}\right)-\left(k+m\right)W_{\alpha}\left(\chi_{\left(k\right)}\psi_{\left(m\right)}\right)\\
 & = & D_{\alpha}\chi_{\left(k\right)}\psi_{\left(m\right)}+\chi_{\left(k\right)}\left(D_{\alpha}\psi_{\left(m\right)}\right)
\end{eqnarray*}

These considerations appy to both scalars and vectors. For weighted,
vector-valued fields $v_{\left(k\right)}^{\beta}$ the covariant derivative
is
\[
D_{\alpha}v_{\left(k\right)}^{\beta}=\nabla_{\alpha}v_{\left(k\right)}^{\beta}+v_{\left(k\right)}^{\mu}\hat{\Gamma}_{\;\;\;\mu\alpha}^{\beta}-kv_{\left(k\right)}^{\beta}W_{\alpha}
\]
where $\hat{\Gamma}_{\;\;\;\mu\alpha}^{\beta}$ provides covariance
under non-abelian transformations and $kW_{\alpha}$ under abelian
transformations. \emph{Notice that a linear transformation $\hat{\Gamma}_{\;\;\;\mu\alpha}^{\beta}dx^{\alpha}$
of dimension $n\geq2$ cannot be restricted to act on scalars}.

Dilatations provide another example of an abelian symmetry. A dilatation
will rescale a dimensionful field such as the volume element $\sqrt{-g}\rightarrow e^{4\varphi}\sqrt{-g}$
when the metric scales as $g_{\alpha\beta}\rightarrow e^{2\varphi}g_{\alpha\beta}$.
The metric is said to be of conformal weight 2, and the volume form
of weight 4, so the scale-covariant derivative of the volume form
is
\[
\hat{D}_{\mu}\sqrt{-g}=\partial_{\mu}\sqrt{-g}-4W_{\mu}\sqrt{-g}
\]
where $W_{\mu}$ is the Weyl vector. Because the metric has nonzero
conformal weight, the general form of the combined general coordinate
and scale covariant derivative is
\begin{equation}
\hat{D}_{\mu}g_{\alpha\beta}=\partial_{\mu}g_{\alpha\beta}-g_{\rho\beta}\hat{\Gamma}_{\;\;\;\alpha\mu}^{\rho}-g_{\alpha\rho}\hat{\Gamma}_{\;\;\;\beta\mu}^{\rho}-2g_{\alpha\beta}W_{\mu}\sqrt{-g}\label{Derivative of the metric}
\end{equation}
This is tensorial under general coordinate transformations with the
usual inhomogeneous transformation of $\hat{\Gamma}_{\;\;\;\alpha\beta}^{\mu}$,
and also under the combined conformal transformation
\begin{eqnarray*}
g_{\alpha\beta} & \rightarrow & e^{2\varphi}g_{\alpha\beta}\\
W_{\mu} & \rightarrow & W_{\mu}+\partial_{\mu}\varphi
\end{eqnarray*}

It is natural to include the possibility of a Weyl geometry when considering
the differential geometry of spacetime. Indeed, other systematic approaches
to the underlying geometry of spacetime also lead to Weyl geometry.
Recent work by Trautman, Matveev, and Scholz \cite{Matveev1,Matveev2}
puts fresh rigor to the physically insightful work of Ehlers, Pirani,
and Schild \cite{EPS}. These studies show that agreement of the projective
structure of timelike geodesics and the conformal structure of lightlike
geodesics in the lightlike limit leads to an integrable Weyl geometry.
Thus, since ultimately we measure only paths of particles, we should
expect the world to be described by a Weyl geometry. For agreement
with experiment it is important that within strong experimental limits
this should be an \emph{integrable Weyl geometry}, in which the Weyl
vector takes the pure gauge form $W_{\mu}=\partial_{\mu}\phi$. There
then exists a gauge in which the Weyl vector vanishes, and transport
of physical objects around closed paths does \emph{not} lead to measurable
relative size change.

Whatever abelian symmetry we envision, when varying the connection
the most general ansatz for the covariant derivative of a weighted
vector is
\[
\hat{D}_{\mu}v_{\left(k\right)}^{\alpha}=\partial_{\mu}v_{\left(k\right)}^{\alpha}+v_{\left(k\right)}^{\beta}\hat{\Gamma}_{\;\;\;\beta\mu}^{\alpha}-kW_{\mu}v_{\left(k\right)}^{\alpha}
\]
Our central point is this: \emph{If the metric has nonzero weight,
then use of Eq.(\ref{Derivative of the metric}) instead of Eq.(\ref{Non-abelian derivative of the metric})
is necessary and will change the result of the Palatini variation.} 

While our discussion applies to any abelian symmetry, our results
apply when the abelian symmetry affects the metric $\left(w_{g}\neq0\right)$.
Given this, it does not matter whether the symmetry is interpreted
as scale covariance or some other physical symmetry. The resulting
structure is always that of a Weyl geometry.

\subsection{The Palatini variation again}

We now repeat the argument of \ref{subsec:The-standard-Palatini}
using the fully general form given in Eq.(\ref{Derivative of the metric})
for the connection. The curvature experienced by a weight zero field
is the of the usual form in terms of $\hat{\Gamma}_{\;\;\;\beta\mu}^{\alpha}$
alone, so the connection variation still takes the form given in Eq.(\ref{Initial connection variation}).
\begin{eqnarray*}
\delta_{\Gamma}S_{P}\left[g,\hat{\Gamma}\right] & = & \int\left(\hat{D}_{\mu}\left(\delta\hat{\Gamma}_{\;\;\;\alpha\beta}^{\mu}\right)-\hat{D}_{\beta}\left(\delta\hat{\Gamma}_{\;\;\;\alpha\mu}^{\mu}\right)\right)g^{\alpha\beta}\sqrt{-g}d^{4}x
\end{eqnarray*}
The only difference is the addition of a possible abelian term in
the derivative of the connection variation,
\[
\hat{D}_{\mu}\left(\delta\hat{\Gamma}_{\;\;\;\alpha\beta}^{\nu}\right)=\nabla_{\mu}\left(\delta\hat{\Gamma}_{\;\;\;\alpha\beta}^{\nu}\right)+\left(\delta\hat{\Gamma}_{\;\;\;\alpha\beta}^{\rho}\right)C_{\;\;\;\rho\mu}^{\nu}-\left(\delta\hat{\Gamma}_{\;\;\;\rho\beta}^{\nu}\right)C_{\;\;\;\alpha\mu}^{\rho}-\left(\delta\hat{\Gamma}_{\;\;\;\alpha\rho}^{\nu}\right)C_{\;\;\;\beta\mu}^{\rho}-w_{\Gamma}W_{\mu}\left(\delta\hat{\Gamma}_{\;\;\;\alpha\beta}^{\nu}\right)
\]
where $w_{\Gamma}$ is the weight of $\delta\hat{\Gamma}_{\;\;\;\alpha\beta}^{\nu}$
and we again separate out the metric compatible $\nabla_{\mu}$. Taking
the required contractions and substituting yields
\begin{eqnarray*}
\delta_{\Gamma}S_{P}\left[g,\hat{\Gamma}\right] & = & \int\left(\nabla_{\mu}\left(\delta\hat{\Gamma}_{\;\;\;\alpha\beta}^{\mu}\right)-\nabla_{\beta}\left(\delta\hat{\Gamma}_{\;\;\;\alpha\nu}^{\nu}\right)\right)g^{\alpha\beta}\sqrt{-g}d^{4}x\\
 &  & +\int\left(\delta\hat{\Gamma}_{\;\;\;\alpha\beta}^{\rho}C_{\;\;\;\rho\mu}^{\mu}-\delta\hat{\Gamma}_{\;\;\;\rho\beta}^{\mu}C_{\;\;\;\alpha\mu}^{\rho}-\delta\hat{\Gamma}_{\;\;\;\alpha\rho}^{\mu}C_{\;\;\;\beta\mu}^{\rho}-w_{\Gamma}W_{\mu}\delta\hat{\Gamma}_{\;\;\;\alpha\beta}^{\mu}\right)g^{\alpha\beta}\sqrt{-g}d^{4}x\\
 &  & +\int\left(\left(\delta\hat{\Gamma}_{\;\;\;\rho\nu}^{\nu}\right)C_{\;\;\;\alpha\beta}^{\rho}+w_{\Gamma}\left(W_{\beta}\delta\hat{\Gamma}_{\;\;\;\alpha\nu}^{\nu}\right)\right)g^{\alpha\beta}\sqrt{-g}d^{4}x
\end{eqnarray*}
Integration by parts of the metric compatible derivative gives zero
acting on $g^{\alpha\beta}\sqrt{-g}$, and we are once again left
with an algebraic condition for the non-metric piece. Factoring out
the variation,

\begin{eqnarray*}
0 & = & \int\left(\delta\hat{\Gamma}_{\;\;\;\sigma\lambda}^{\rho}\left(\delta_{\alpha}^{\sigma}\delta_{\beta}^{\lambda}C_{\;\;\;\rho\mu}^{\mu}-\delta_{\beta}^{\lambda}C_{\;\;\;\alpha\rho}^{\sigma}-\delta_{\alpha}^{\sigma}C_{\;\;\;\beta\rho}^{\lambda}-w_{\Gamma}W_{\rho}\delta_{\alpha}^{\sigma}\delta_{\beta}^{\lambda}\right)\right)g^{\alpha\beta}\sqrt{-g}d^{4}x\\
 &  & +\int\delta\hat{\Gamma}_{\;\;\;\sigma\lambda}^{\rho}\left(\delta_{\rho}^{\lambda}C_{\;\;\;\alpha\beta}^{\sigma}+w_{\Gamma}W_{\beta}\delta_{\alpha}^{\sigma}\delta_{\rho}^{\lambda}\right)g^{\alpha\beta}\sqrt{-g}d^{4}x
\end{eqnarray*}
and carrying out the contractions with the metric the field equation
becomes
\begin{eqnarray}
0 & = & g^{\sigma\lambda}C_{\;\;\;\rho\mu}^{\mu}-C_{\quad\rho}^{\sigma\lambda}-C_{\quad\rho}^{\lambda\sigma}-w_{\Gamma}W_{\rho}g^{\sigma\lambda}+\delta_{\rho}^{\lambda}C_{\quad\beta}^{\sigma\beta}+w_{\Gamma}W^{\sigma}\delta_{\rho}^{\lambda}\label{New field equation}
\end{eqnarray}
Now the $\sigma\lambda$ contraction becomes
\begin{eqnarray*}
0 & = & 2C_{\;\;\;\rho\mu}^{\mu}+C_{\rho\quad\;\;\beta}^{\;\;\;\beta}-3w_{\Gamma}W_{\rho}
\end{eqnarray*}
The $\sigma\rho$ trace vanishes identically, while contracting $\lambda\rho$
gives
\begin{eqnarray*}
0 & = & 3C_{\quad\beta}^{\sigma\beta}+3w_{\Gamma}W^{\sigma}
\end{eqnarray*}
and therefore, solving we have
\begin{eqnarray*}
C_{\quad\beta}^{\sigma\beta} & = & -w_{\Gamma}W^{\sigma}\\
C_{\;\;\;\rho\mu}^{\mu} & = & 2w_{\Gamma}W_{\rho}
\end{eqnarray*}

Substituting the contractions back into Eq.(\ref{New field equation}),
\begin{eqnarray*}
0 & = & 2w_{\Gamma}W_{\rho}g^{\sigma\lambda}-C_{\quad\rho}^{\sigma\lambda}-C_{\quad\rho}^{\lambda\sigma}-w_{\Gamma}W_{\rho}g^{\sigma\lambda}-\delta_{\rho}^{\lambda}w_{\Gamma}W^{\sigma}+w_{\Gamma}W^{\sigma}\delta_{\rho}^{\lambda}\\
C_{\quad\rho}^{\sigma\lambda}+C_{\quad\rho}^{\lambda\sigma} & = & w_{\Gamma}W_{\rho}g^{\sigma\lambda}
\end{eqnarray*}
Lowering the indices, we permute indices and combine in the usual
way to isolate $C_{\lambda\sigma\rho}$.
\begin{eqnarray*}
C_{\sigma\lambda\rho}+C_{\lambda\sigma\rho}+C_{\lambda\rho\sigma}+C_{\rho\lambda\sigma}-C_{\rho\sigma\lambda}-C_{\sigma\rho\lambda} & = & w_{\Gamma}W_{\rho}g_{\sigma\lambda}+w_{\Gamma}W_{\sigma}g_{\lambda\rho}-w_{\Gamma}W_{\lambda}g_{\rho\sigma}\\
C_{\lambda\rho\sigma} & = & \frac{1}{2}\left(w_{\Gamma}W_{\rho}g_{\sigma\lambda}+w_{\Gamma}W_{\sigma}g_{\lambda\rho}-w_{\Gamma}W_{\lambda}g_{\rho\sigma}\right)
\end{eqnarray*}
Restoring the first index to its natural position
\begin{eqnarray*}
C_{\;\;\;\sigma\rho}^{\lambda} & = & \frac{1}{2}w_{\Gamma}\left(\delta_{\sigma}^{\lambda}W_{\rho}+\delta_{\rho}^{\lambda}W_{\sigma}-W^{\lambda}g_{\rho\sigma}\right)
\end{eqnarray*}

We can choose the weight $w_{\Gamma}$ to insure metric compatibility.
With this expression for $C_{\;\;\;\sigma\rho}^{\lambda}$, the covariant
derivative of a weight $w_{g}$ metric is 
\begin{eqnarray*}
\hat{D}_{\rho}g_{\alpha\beta} & = & g_{\alpha\beta,\rho}-g_{\lambda\beta}\hat{\Gamma}_{\;\;\;\alpha\rho}^{\lambda}-g_{\alpha\lambda}\hat{\Gamma}_{\;\;\;\beta\rho}^{\lambda}-w_{g}W_{\rho}g_{\alpha\beta}\\
 & = & g_{\alpha\beta,\rho}-g_{\lambda\beta}\left(\frac{1}{2}g^{\lambda\mu}\left(g_{\mu\alpha,\rho}+g_{\mu\rho,\alpha}-g_{\alpha\rho,\mu}\right)+\frac{1}{2}w_{\Gamma}g^{\lambda\mu}\left(g_{\mu\alpha}W_{\rho}+g_{\mu\rho}W_{\alpha}-g_{\rho\alpha}W_{\mu}\right)\right)\\
 &  & -g_{\alpha\lambda}\left(\frac{1}{2}g^{\lambda\mu}\left(g_{\mu\beta,\rho}+g_{\mu\rho,\beta}-g_{\beta\rho,\mu}\right)+\frac{1}{2}w_{\Gamma}g^{\lambda\mu}\left(g_{\mu\beta}W_{\rho}+g_{\mu\rho}W_{\beta}-g_{\rho\beta}W_{\mu}\right)\right)-w_{g}W_{\rho}g_{\alpha\beta}\\
 &  & -\frac{1}{2}w_{g}g_{\beta\alpha}W_{\rho}-\frac{1}{2}w_{g}g_{\beta\rho}W_{\alpha}+\frac{1}{2}w_{g}g_{\rho\alpha}W_{\beta}-\frac{1}{2}w_{g}g_{\alpha\beta}W_{\rho}-\frac{1}{2}w_{g}g_{\alpha\rho}W_{\beta}+\frac{1}{2}w_{g}g_{\rho\beta}W_{\alpha}-2W_{\rho}g_{\alpha\beta}\\
 & = & -w_{\Gamma}g_{\beta\alpha}W_{\rho}-w_{g}W_{\rho}g_{\alpha\beta}
\end{eqnarray*}
so we set $w_{\Gamma}=-w_{g}$. Then despite the non-vanishing of
$C_{\;\;\;\sigma\rho}^{\lambda}$ we have metric compatibity. The
full connection is
\begin{eqnarray}
\hat{\Gamma}_{\;\;\;\alpha\beta}^{\nu} & = & \frac{1}{2}g^{\nu\mu}\left(g_{\mu\alpha,\rho}+g_{\mu\rho,\alpha}-g_{\alpha\rho,\mu}\right)-\frac{w_{g}}{2}g^{\nu\mu}\left(g_{\mu\alpha}W_{\rho}+g_{\mu\rho}W_{\alpha}-g_{\rho\alpha}W_{\mu}\right)\nonumber \\
 & = & \frac{1}{2}g^{\nu\mu}\left(\mathcal{D}_{\rho}g_{\mu\alpha}+\mathcal{D}_{\alpha}g_{\mu\rho}-\mathcal{D}_{\mu}g_{\alpha\rho}\right)\label{Weyl connection}
\end{eqnarray}
where
\[
\mathcal{D}_{\mu}g_{\alpha\beta}\equiv g_{\alpha\beta,\mu}-w_{g}g_{\alpha\beta}W_{\mu}
\]
is the abelian-covariant derivative of the metric. This makes the
full connection invariant under the abelian transformations. The $W_{\alpha}$
terms in Eq.(\ref{Weyl connection}) represent \emph{decoupling} of
the abelian and non-abelian parts of the derivative.

Equation(\ref{Weyl connection}) is the connection of a Weyl geometry.
In this sense, the Palatini variation leads to a Weyl geometry.

\subsection{Integrability of the Weyl geometry}

When the weight of the metric is nonzero, $w_{g}\neq0$, the defining
vector of a Weyl geometry, $W_{\mu}$, is called the Weyl vector.
Through its coupling to the metric it affects lengths. Suppose $s^{\alpha}$
is a constant, weight zero vector associated with a physical object
so that in a flat geometry ($\Gamma_{\;\;\;\mu\nu}^{\alpha}=0$) we
have 
\[
\mathcal{D}_{\alpha}s^{\beta}=\partial_{\alpha}s^{\beta}-0\cdot W_{\alpha}s^{\alpha}=0
\]
Then the covariant derivative of $s^{2}=\eta_{\alpha\beta}s^{\alpha}s^{\beta}$
is
\begin{eqnarray*}
\mathcal{D}_{\mu}s^{2} & = & \mathcal{D}_{\mu}\left(\eta_{\alpha\beta}s^{\alpha}s^{\beta}\right)\\
 & = & \left(\mathcal{D}_{\mu}\eta_{\alpha\beta}\right)s^{\alpha}s^{\beta}\\
 & = & -w_{g}W_{\mu}s^{a}
\end{eqnarray*}
If two identical such rods are carried along different paths and brought
back together forming a closed curve $C$, their lengths no longer
match but differ by
\[
\Delta s^{2}=-w_{g}\ointop_{C}W_{\mu}dx^{\mu}=-w_{g}\iintop_{S}\left(\partial_{\mu}W_{\nu}-\partial_{\nu}W_{\mu}\right)dS^{\mu\nu}
\]
This constitutes a measurable change in physical size unless the curl
of the Weyl vector vanishes. Even on small scales such an effect would
drastically spread atomic, nuclear, and particle spectral lines and
resonances, in conflict with experiment. Therefore, it is important
that unless the coupling is immeasurably small, the curl of the Weyl
vector must vanish. When this is the case, the Weyl connection describes
an \emph{integrable Weyl geometry}. In an integrable Weyl geometry
$W_{\mu}$ is a gradient, and there exists a rescaling such that $\tilde{W}_{\mu}=0$,
returning the appearance of the geometry to Riemannian\footnote{Even in the \emph{Riemannian gauge} with $W_{\mu}=0$, there is still
a difference between a Weyl geometry and a Riemannian geometry, since
in the former a rescaling will restore a nonzero Weyl vector while
keeping physical scalars unchanged, while the same rescaling will
substantially change physical predictions in a Riemannian geometry. }.

With this background in mind, we examine the form of the Weyl vector
by looking in detail at the derivative of the volume form, $g=\det\left(g_{\alpha\beta}\right)$.

For the abelian symmetry we know that
\begin{eqnarray}
\hat{D}_{\mu}g & = & \partial_{\mu}g-4w_{g}gW_{\mu}\label{Derivative of weighted volume}
\end{eqnarray}
Expanding the determinant in terms of the metric,
\[
g=\det g_{\alpha\beta}=\frac{1}{4!}\varepsilon^{\alpha\beta\mu\nu}\varepsilon^{\rho\sigma\lambda\tau}g_{\alpha\rho}g_{\beta\sigma}g_{\mu\lambda}g_{\nu\tau}
\]
we may express $\hat{D}_{\mu}g$ in terms of the metric derivative,

\begin{eqnarray}
\hat{D}_{\mu}g & = & \frac{1}{3!}\varepsilon^{\alpha\beta\varphi\nu}\varepsilon^{\rho\sigma\lambda\tau}\left(\hat{D}_{\mu}g_{\alpha\rho}\right)g_{\beta\sigma}g_{\varphi\lambda}g_{\nu\tau}=\Sigma^{\alpha\rho}\left(\hat{D}_{\mu}g_{\alpha\rho}\right)\label{Dhat g}
\end{eqnarray}
We define
\begin{eqnarray*}
\Sigma^{\alpha\rho} & \equiv & \frac{1}{3!}\varepsilon^{\alpha\beta\varphi\nu}\varepsilon^{\rho\sigma\lambda\tau}g_{\beta\sigma}g_{\varphi\lambda}g_{\nu\tau}\\
 & = & -\frac{1}{3!}ge^{\alpha\beta\varphi\nu}e^{\rho\sigma\lambda\tau}g_{\beta\sigma}g_{\varphi\lambda}g_{\nu\tau}
\end{eqnarray*}
where $e^{\alpha\beta\varphi\nu}=\frac{1}{\sqrt{-g}}\varepsilon^{\alpha\beta\varphi\nu}$
is the Levi-Civita tensor. Then contracting with another copy of the
metric, we lower the indices on the second Levi-Civita tensor,
\begin{eqnarray*}
\Sigma^{\alpha\rho}g_{\rho\theta} & = & -\frac{1}{3!}ge^{\alpha\beta\varphi\nu}e^{\rho\sigma\lambda\tau}g_{\rho\theta}g_{\beta\sigma}g_{\varphi\lambda}g_{\nu\tau}\\
 & = & -\frac{1}{3!}ge^{\alpha\beta\varphi\nu}e_{\theta\beta\varphi\nu}\\
 & = & g\delta_{\theta}^{\alpha}
\end{eqnarray*}
This shows that
\[
\Sigma^{\alpha\rho}=gg^{\alpha\rho}
\]
since the inverse metric is unique and the volume element nonvanishing.
Therefore, returning to Eq.(\ref{Dhat g}), 
\[
\hat{D}_{\mu}g=gg^{\alpha\rho}\hat{D}_{\mu}g_{\alpha\rho}
\]
The same argument shows that the partial derivative of the metric
determinant is
\begin{equation}
\partial_{\mu}g=gg^{\alpha\rho}\partial_{\mu}g_{\alpha\rho}\label{Partial of the volume form}
\end{equation}

With the covariant derivative of the metric given by
\begin{eqnarray*}
\hat{D}_{\mu}g_{\alpha\beta} & = & \partial_{\mu}g_{\alpha\beta}-g_{\nu\beta}\hat{\Gamma}_{\;\;\;\alpha\mu}^{\nu}-g_{\alpha\nu}\hat{\Gamma}_{\;\;\;\beta\mu}^{\nu}-w_{g}W_{\mu}g_{\alpha\beta}
\end{eqnarray*}
the covariant derivative of the volume form becomes
\begin{eqnarray}
\hat{D}_{\mu}g & = & gg^{\alpha\beta}\hat{D}_{\mu}g_{\alpha\beta}\nonumber \\
 & = & gg^{\alpha\beta}\partial_{\mu}g_{\alpha\beta}-2g\hat{\Gamma}_{\;\;\;\alpha\mu}^{\alpha}-4w_{g}gW_{\mu}\label{Cov deriv of g}
\end{eqnarray}
Substituting Eqs.(\ref{Partial of the volume form}) into (\ref{Cov deriv of g})
and equating to the expression in Eq.(\ref{Derivative of weighted volume}),
\[
\partial_{\mu}g-4w_{g}gW_{\mu}=\partial_{\mu}g-2g\hat{\Gamma}_{\;\;\;\alpha\mu}^{\alpha}-4w_{g}gW_{\mu}
\]
and therefore
\begin{eqnarray*}
\hat{\Gamma}_{\;\;\;\alpha\mu}^{\alpha} & = & 0
\end{eqnarray*}
From the form of the Weyl connection Eq.(\ref{Weyl connection}) this
implies
\begin{eqnarray*}
\hat{\Gamma}_{\;\;\;\alpha\beta}^{\alpha} & = & \frac{1}{2}g^{\alpha\mu}\left(g_{\mu\alpha,\beta}+g_{\mu\beta,\alpha}-g_{\alpha\beta,\mu}\right)-\frac{w_{g}}{2}g^{\alpha\mu}\left(g_{\mu\alpha}W_{\beta}+g_{\mu\beta}W_{\alpha}-g_{\beta\alpha}W_{\mu}\right)\\
0 & = & \frac{1}{2}g^{\alpha\mu}g_{\mu\alpha,\beta}-\frac{w_{g}}{2}\left(4W_{\beta}+W_{\beta}-W_{\beta}\right)\\
0 & = & \frac{1}{2}g^{\alpha\mu}g_{\mu\alpha,\beta}-2w_{g}W_{\beta}
\end{eqnarray*}
and therefore
\begin{eqnarray*}
W_{\beta} & = & \frac{1}{4w_{g}}g^{\alpha\mu}g_{\mu\alpha,\beta}\\
 & = & \frac{1}{4w_{g}}\partial_{\mu}\left(\ln g\right)
\end{eqnarray*}
The Weyl vector is therefore a gradient and the Palatini variation
leads to an \emph{integrable Weyl geometry}. This is in good agreement
with experiment and with the conclusions about measurability by Matveev
and Trautman \cite{Matveev1}.

The result is striking in two ways. First, the integrability of the
Weyl vector means that there exists a choice of gauge in which the
field equation takes the usual form from general relativity. In this
sense, the usual Palatini conclusion holds: the Palatini action $S\left[g,\hat{\Gamma},W\right]$
leads to the usual Einstein equation together with the Christoffel
connection, \emph{but only in a particular conformal gauge}.

The second striking feature is that we are led by the Palatini variation
to an integrable Weyl geometry. In this sense, the usual conclusion
is wrong. We do not get only the Christoffel connection. The physical
arguments of \cite{EPS,Matveev1} are supported by the free variation
of the connection.

Before concluding, we must ask whether the presence of sources will
make the Weyl vector non-integrable. To high precision this would
conflict with observation. However, the Einstein tensor of a Weyl
geometry is given by \cite{Wheeler}
\begin{eqnarray*}
\mathfrak{G}_{ab} & = & R_{ab}-\frac{1}{2}R\eta_{ab}+2W_{a;b}+2W_{a}W_{b}+\left(W^{2}-2W_{\;\;\;;c}^{c}\right)\eta_{ab}
\end{eqnarray*}
and this will equal the energy tensor of the source, which in turn
arises from the metric variation of the matter action. Since the metric
is symmetric, this always yields a symmetric source tensor,
\begin{eqnarray*}
\mathfrak{G}_{ab} & = & \kappa T_{ab}
\end{eqnarray*}
Taking the antisymmetric part of this expression leaves only one term,
\[
0=\mathfrak{G}_{\left[ab\right]}=W_{\left[a;b\right]}
\]
and this is the condition for the Weyl vector to be pure gauge. Therefore,
even including sources, Palatini variation leads to an integrable
Weyl geometry and is therefore gauge-equivalent to general relativity.

It is amusing to note that this conclusion agrees with the result
of Einstein in his original formulation of the Palatini variation
\cite{Einstein}. In \cite{Einstein} the metric is replaced by an
asymmetric tensor density and the connection is fully general. The
variation leads to the introduction of a vector field in addition
to the usual metric and connection, even when the metric is symmetric.
Einstein's full argument is presented in the Appendix.

\section{Different independent variables}

When we write general relativity as a Poincarè gauge theory the form
of the metric and connection are altered to give a local Lorentz fiber
bundle. The change of variables begins by replacing coordinate differentials
by an orthonormal 1-form basis. Whereas the metric in a general coordinate
basis is related to coordinate 1-form basis by
\begin{eqnarray*}
\left\langle \mathbf{d}x^{\alpha},\mathbf{d}x^{\beta}\right\rangle  & = & g^{\alpha\beta}
\end{eqnarray*}
the solder form is an orthonormal linear combination $\mathbf{e}^{a}=e_{\alpha}^{\;\;\;a}\mathbf{d}x^{\alpha}$
such that
\[
\left\langle \mathbf{e}^{a},\mathbf{e}^{b}\right\rangle =\eta^{ab}
\]
where $\eta_{ab}$ is the Minkowski metric. Preserving the orthonormality
of the frame field reduces the symmetry from local general linear
to local Lorentz while maintaining complete generality of the geometry.
The change replaces the coordinate metric and connection $\left(g,\hat{\Gamma}\right)$
of the Palatini action with the solder form and spin connection, $\left(\mathbf{e}^{a},\boldsymbol{\omega}_{\;\;\;b}^{a}\right)$.

The Cartan structure equations take the form
\begin{eqnarray}
\mathbf{d}\boldsymbol{\omega}_{\;\;\;b}^{a} & = & \boldsymbol{\omega}_{\;\;\;b}^{c}\land\boldsymbol{\omega}_{\;\;\;c}^{a}+\boldsymbol{\mathcal{R}}_{\;\;\;b}^{a}\label{Curvature}\\
\mathbf{d}\mathbf{e}^{a} & = & \mathbf{e}^{b}\land\boldsymbol{\omega}_{\;\;\;b}^{a}+\mathbf{T}^{a}\label{Torsion}
\end{eqnarray}
where $\boldsymbol{\mathcal{R}}_{\;\;\;b}^{a}$ is the curvature 2-form
and $\mathbf{T}^{a}$ is the torsion 2-form. These require integrability
conditions (Bianchi identities) similar to general relativity,
\begin{eqnarray*}
\mathbf{D}\mathbf{T}^{a} & = & \mathbf{e}^{b}\wedge\boldsymbol{\mathcal{R}}_{\;\;\;b}^{a}\\
\mathbf{D}\boldsymbol{\mathcal{R}}_{\;\;\;b}^{a} & = & 0
\end{eqnarray*}
but the first Bianchi identity now involves the torsion.

To achieve the Riemannian geometry of general relativity directly
we would set the torsion to zero. This eliminates torsion dependence
of the curvature, $\boldsymbol{\mathcal{R}}_{\;\;\;b}^{a}\rightarrow\mathbf{R}_{\;\;\;b}^{a}$.
Then, along with the correspondingly reduced Bianchi identity $0=\mathbf{e}^{b}\wedge\mathbf{R}_{\;\;\;b}^{a}$,
the reduced structure equations
\begin{eqnarray*}
\mathbf{d}\boldsymbol{\omega}_{\;\;\;b}^{a} & = & \boldsymbol{\omega}_{\;\;\;b}^{c}\land\boldsymbol{\omega}_{\;\;\;c}^{a}+\mathbf{R}_{\;\;\;b}^{a}\\
\mathbf{d}\mathbf{e}^{a} & = & \mathbf{e}^{b}\land\boldsymbol{\omega}_{\;\;\;b}^{a}
\end{eqnarray*}
describe a Riemannian geometry. Solder form or metric variation of
the Einstein-Hilbert action
\[
S_{EH}\left[\mathbf{e}^{a}\right]=\frac{1}{2}\int\mathbf{R}^{ab}\wedge\mathbf{e}^{c}\wedge\mathbf{e}^{d}e_{abcd}
\]
gives the Einstein equation. Varying the solder form alone, the torsion
makes no appearance in the field equations, so setting torsion to
zero is consistent throughout.

When torsion is not set to zero by hand, the variation of the Einstein-Hilbert
action with respect to the solder form or metric leads to the Einstein-Cartan-Sciama-Kibble
(ECSK) theory of gravity. While vacuum ECSK theory still leads to
vanishing torsion, torsion can be nonzero in the presence of spinor
sources.

By contrast, the Palatini variation introduces a second field equation
directly dependent on the torsion. While some variants of ECSK theory
vary the metric and torsion, in the gauge theory formulation it is
natural to take the Cartan connection 1-forms as the independent variables.
For Poincarè gauge theory these are the solder form $\mathbf{e}^{a}$
and the spin connection $\boldsymbol{\omega}_{\;\;\;b}^{a}$.

We study whether this change in the choice of independent variables
affects our conclusions regarding the Palatini variation.

Retaining the Einstein-Hilbert action, we write it as
\[
S_{P}\left[\mathbf{e}^{a},\boldsymbol{\omega}_{\;\;\;b}^{a}\right]=\frac{1}{2}\int\boldsymbol{\mathcal{R}}^{ab}\wedge\mathbf{e}^{c}\wedge\mathbf{e}^{d}e_{abcd}
\]
with the curvature and connection given by Eqs.(\ref{Curvature})
and (\ref{Torsion}). Working within the rigid context of these Cartan
structure equations, there is no freedom to modify the connection
as in the previous Section.

\subsection{Solder form variation}

There are no surprises when we vary the solder form.
\begin{eqnarray*}
\delta_{e}S_{P} & = & \int\boldsymbol{\mathcal{R}}^{ab}\wedge\delta\mathbf{e}^{c}\wedge\mathbf{e}^{d}e_{abcd}\\
 & = & \int\boldsymbol{\mathcal{R}}^{ab}\wedge e_{e}^{\;\;\;\mu}\delta e_{\mu}^{\;\;\;c}\mathbf{e}^{e}\wedge\mathbf{e}^{d}e_{abcd}
\end{eqnarray*}
Defining a volume form as the dual of one, $\boldsymbol{\Phi}={}^{*}1=\frac{1}{4!}\varepsilon_{abcd}\,\mathbf{e}^{a}\land\mathbf{e}^{b}\land\mathbf{e}^{c}\land\mathbf{e}^{d}$,
so that
\begin{eqnarray*}
\mathbf{e}^{a}\land\mathbf{e}^{b}\land\mathbf{e}^{c}\land\mathbf{e}^{d} & = & -e^{abcd}\boldsymbol{\Phi}
\end{eqnarray*}
the field equation becomes
\begin{eqnarray*}
0 & = & \boldsymbol{\mathcal{R}}^{ab}\wedge\mathbf{e}^{e}\wedge\mathbf{e}^{d}e_{abcd}\\
 & = & -\frac{1}{2}\mathcal{R}_{\quad fg}^{ab}e^{fged}e_{abcd}\boldsymbol{\Phi}
\end{eqnarray*}
and reducing the doubled Levi-Civita tensor $e^{fgde}e_{abdc}=-\delta_{a}^{[f}\delta_{b}^{g}\delta_{c}^{e]}$
we have the Einstein equation
\begin{eqnarray*}
\mathcal{R}_{ab}-\frac{1}{2}\mathcal{R}\eta_{ab} & = & 0
\end{eqnarray*}
The only difference is that the curvature is that of an Einstein-Cartan
geometry, hence dependent upon the torsion.

\subsection{Varying the spin connection}

Varying the spin connection, some features emerge as before and some
are different. All of the structure is determined by the Cartan equations,
Eqs.(\ref{Curvature}) and (\ref{Torsion}). Using Eq.(\ref{Curvature})
we have

\begin{eqnarray*}
\delta_{\omega}S_{P} & = & \frac{1}{2}\int\mathbf{D}\left(\delta\boldsymbol{\omega}^{ab}\right)\wedge\mathbf{e}^{c}\wedge\mathbf{e}^{d}e_{abcd}
\end{eqnarray*}
where 
\begin{eqnarray*}
\mathbf{D}\delta\boldsymbol{\omega}^{ab} & = & \mathbf{d}\left(\delta\boldsymbol{\omega}^{ab}\right)+\boldsymbol{\omega}_{\;\;\;e}^{a}\land\delta\boldsymbol{\omega}_{\;\;\;b}^{e}-\boldsymbol{\omega}_{\;\;\;b}^{e}\land\delta\boldsymbol{\omega}_{\;\;\;e}^{a}
\end{eqnarray*}
Integrating by parts we need to exercise caution because $\frac{1}{2}\int\mathbf{D}\left(\delta\boldsymbol{\omega}^{ab}\wedge\mathbf{e}^{c}\wedge\mathbf{e}^{d}e_{abcd}\right)$
is not necessarily just a surface term. From the Leibnitz rule we
must have
\begin{eqnarray*}
\delta_{\omega}S_{P} & = & \frac{1}{2}\int\mathbf{D}\left(\delta\boldsymbol{\omega}^{ab}\wedge\mathbf{e}^{c}\wedge\mathbf{e}^{d}e_{abcd}\right)+\frac{1}{2}\int\delta\boldsymbol{\omega}^{ab}\wedge\mathbf{D}\left(\mathbf{e}^{c}\wedge\mathbf{e}^{d}e_{abcd}\right)
\end{eqnarray*}
but if there is a non-abelian symmetry the action of $\mathbf{D}$
on a scalar is not just the exterior derivative. Rather,
\[
\mathbf{D}\left(\delta\boldsymbol{\omega}^{ab}\wedge\mathbf{e}^{c}\wedge\mathbf{e}^{d}e_{abcd}\right)=\mathbf{d}\left(\delta\boldsymbol{\omega}^{ab}\wedge\mathbf{e}^{c}\wedge\mathbf{e}^{d}e_{abcd}\right)-w_{\Sigma}\,\boldsymbol{\omega}\land\left(\delta\boldsymbol{\omega}^{ab}\wedge\mathbf{e}^{c}\wedge\mathbf{e}^{d}e_{abcd}\right)
\]
where $w_{\Sigma}$ is the weight of the scalar 3-form
\begin{eqnarray*}
\boldsymbol{\Sigma} & \equiv & \delta\boldsymbol{\omega}^{ab}\wedge\mathbf{e}^{c}\wedge\mathbf{e}^{d}e_{abcd}
\end{eqnarray*}

For the action to have an abelian symmetry, its weight should be zero.
Given the weight $w_{g}$ of the metric, it is straightforward to
determine the weights of the remaining fields. This is carried out
for all relevant fields in Appendix II to show that
\[
\begin{array}{ccccccc}
w\left(\mathbf{e}^{a}\right) & = & \frac{1}{2}w_{g} &  & w\left(\boldsymbol{\omega}_{\;\;\;b}^{a}\right) & = & 0\\
w\left(\boldsymbol{\Phi}\right) & = & 2w_{g} &  & w\left(\omega_{\;\;\;bc}^{a}\right) & = & -\frac{1}{2}w_{g}\\
w\left(e_{abcd}\right) & = & 0 &  & w\left(\mathbf{T}^{a}\right) & = & \frac{1}{2}w_{g}\\
w\left(\eta_{ab}\right) & = & 0 &  & w_{\Sigma} & = & w_{g}
\end{array}
\]
In particular we have $w_{\Sigma}=w_{g}$.

Returning to the variation
\begin{eqnarray*}
\delta_{\omega}S_{P} & = & \frac{1}{2}\int\mathbf{D}\left(\delta\boldsymbol{\omega}^{ab}\wedge\mathbf{e}^{c}\wedge\mathbf{e}^{d}e_{abcd}\right)+\frac{1}{2}\int\delta\boldsymbol{\omega}^{ab}\wedge\mathbf{D}\left(\mathbf{e}^{c}\wedge\mathbf{e}^{d}e_{abcd}\right)\\
 & = & \frac{1}{2}\int\mathbf{d}\left(\delta\boldsymbol{\omega}^{ab}\wedge\mathbf{e}^{c}\wedge\mathbf{e}^{d}e_{abcd}\right)-\frac{1}{2}\int w_{\Sigma}\boldsymbol{\omega}\land\left(\delta\boldsymbol{\omega}^{ab}\wedge\mathbf{e}^{c}\wedge\mathbf{e}^{d}e_{abcd}\right)+\frac{1}{2}\int\delta\boldsymbol{\omega}^{ab}\wedge\mathbf{D}\left(\mathbf{e}^{c}\wedge\mathbf{e}^{d}e_{abcd}\right)
\end{eqnarray*}
Discarding the surface term, writing $\delta\boldsymbol{\omega}^{ab}=\delta\omega_{\quad e}^{ab}\mathbf{e}^{e}$,
then setting the variation to zero, the field equation becomes
\begin{eqnarray*}
0 & = & \frac{1}{2}\mathbf{e}^{e}\wedge\mathbf{D}\left(\mathbf{e}^{c}\wedge\mathbf{e}^{d}e_{abcd}\right)+\frac{1}{2}w_{\Sigma}\mathbf{e}^{e}\wedge\boldsymbol{\omega}\land\mathbf{e}^{c}\wedge\mathbf{e}^{d}e_{abcd}\\
 & = & \frac{1}{2}\mathbf{e}^{e}\wedge\left(\mathbf{D}\mathbf{e}^{c}\wedge\mathbf{e}^{d}e_{abcd}-\mathbf{e}^{c}\wedge\mathbf{D}\mathbf{e}^{d}e_{abcd}+\mathbf{e}^{c}\wedge\mathbf{e}^{d}\mathbf{D}e_{abcd}\right)+\frac{1}{2}w_{g}W_{f}\mathbf{e}^{e}\wedge\mathbf{e}^{f}\land\mathbf{e}^{c}\wedge\mathbf{e}^{d}e_{abcd}\\
 & = & \mathbf{T}^{c}\wedge\mathbf{e}^{e}\wedge\mathbf{e}^{d}e_{abcd}+\frac{1}{2}\mathbf{e}^{e}\wedge\mathbf{e}^{c}\wedge\mathbf{e}^{d}\mathbf{D}e_{abcd}-\frac{1}{2}w_{g}W_{f}e^{efcd}e_{abcd}\boldsymbol{\Phi}
\end{eqnarray*}
and since $w\left(e_{abcd}\right)=0$,
\begin{eqnarray*}
\mathbf{D}e_{abcd} & = & \mathbf{d}e_{abcd}=0
\end{eqnarray*}
We are left with
\begin{eqnarray*}
0 & = & \frac{1}{2}T_{\;\;\;fg}^{c}\mathbf{e}^{f}\wedge\mathbf{e}^{g}\wedge\mathbf{e}^{e}\wedge\mathbf{e}^{d}e_{abcd}+w_{g}W_{f}\left(\delta_{a}^{e}\delta_{b}^{f}-\delta_{a}^{f}\delta_{b}^{e}\right)\boldsymbol{\Phi}\\
0 & = & -T_{\;\;\;fg}^{c}e^{efgd}e_{abcd}+w_{g}\left(\delta_{a}^{e}W_{b}-\delta_{b}^{e}W_{a}\right)
\end{eqnarray*}
Resolving the double Levi-Civita, the field equation becomes
\begin{eqnarray*}
0 & = & T_{\;\;\;ab}^{e}+T_{\;\;\;da}^{d}\delta_{b}^{e}-T_{\;\;\;db}^{d}\delta_{a}^{e}+w_{g}\left(\delta_{a}^{e}W_{b}-\delta_{b}^{e}W_{a}\right)
\end{eqnarray*}

The $ea$ trace of the field equation shows that
\begin{eqnarray*}
T_{\;\;\;ab}^{a} & = & \frac{3}{2}w_{g}W_{b}
\end{eqnarray*}
so that
\begin{eqnarray*}
T_{\;\;\;ab}^{e} & = & \frac{1}{2}w_{g}\left(\delta_{a}^{e}W_{b}-\delta_{b}^{e}W_{a}\right)
\end{eqnarray*}
Writing the torsion as a 2-form,
\begin{eqnarray*}
\mathbf{T}^{a} & = & \frac{1}{2}w_{g}\mathbf{e}^{a}\land\boldsymbol{\omega}
\end{eqnarray*}

As a Riemannian geometry, the spacetime has torsion. However, substituting
into the Cartan structure equations yields
\begin{eqnarray}
\mathbf{d}\boldsymbol{\omega}_{\;\;\;b}^{a} & = & \boldsymbol{\omega}_{\;\;\;b}^{c}\land\boldsymbol{\omega}_{\;\;\;c}^{a}+\boldsymbol{\mathcal{R}}_{\;\;\;b}^{a}\nonumber \\
\mathbf{d}\mathbf{e}^{a} & = & \mathbf{e}^{b}\land\boldsymbol{\omega}_{\;\;\;b}^{a}+\frac{1}{2}w_{g}\mathbf{e}^{a}\land\boldsymbol{\omega}\label{Weyl structure eqs}
\end{eqnarray}
These are the structure equations of Weyl geometry \cite{Wheeler}.
Since $w_{e}=\frac{1}{2}w_{g}$ this shows that the Weyl covariant
derivative of the solder form vanishes,
\begin{eqnarray*}
\mathbf{D}\mathbf{e}^{a} & = & \mathbf{d}\mathbf{e}^{a}-\mathbf{e}^{b}\land\boldsymbol{\omega}_{\;\;\;b}^{a}-w_{e}\mathbf{e}^{a}\land\boldsymbol{\omega}=0
\end{eqnarray*}
Taking Eqs.(\ref{Weyl structure eqs}) as the structure equations
of a Weyl geometry, the Weyl connection is metric compatible and torsion
free.

\subsection{Integrability of the Weyl geometry}

With the structure equations now in the form
\begin{eqnarray}
\mathbf{d}\boldsymbol{\omega}_{\;\;\;b}^{a} & = & \boldsymbol{\omega}_{\;\;\;b}^{c}\land\boldsymbol{\omega}_{\;\;\;c}^{a}+\boldsymbol{\mathcal{R}}_{\;\;\;b}^{a}\label{Curvature of Weyl geometry}\\
\mathbf{d}\mathbf{e}^{a} & = & \mathbf{e}^{b}\land\boldsymbol{\omega}_{\;\;\;b}^{a}+w_{e}\mathbf{e}^{a}\land\boldsymbol{\omega}\label{Structure eq of Weyl geometry}
\end{eqnarray}
we may find the contribution of $\boldsymbol{\omega}$ to the curvature.
Observing that the solution to Eq.(\ref{Structure eq of Weyl geometry})
for the spin connection must be the Riemannian spin connection plus
a term linear in $W_{a}$, we set
\[
\boldsymbol{\omega}_{\;\;\;b}^{a}=\boldsymbol{\alpha}_{\;\;\;b}^{a}+\boldsymbol{\beta}_{\;\;\;b}^{a}
\]
where
\[
\mathbf{d}\mathbf{e}^{a}=\mathbf{e}^{b}\land\boldsymbol{\alpha}_{\;\;\;b}^{a}
\]
defines the Christoffel spin connection and we let
\[
\boldsymbol{\beta}_{\;\;\;b}^{a}=\alpha\eta_{bc}\mathbf{e}^{c}W^{a}+\beta\mathbf{e}^{a}W_{b}
\]
Antisymmetry on $ab$ requires $\alpha=-\beta$. Substituting into
Eq.(\ref{Structure eq of Weyl geometry}) leaves $\beta\mathbf{e}^{a}\land\boldsymbol{\omega}=\mathbf{e}^{a}\land\boldsymbol{\omega}$
so $\beta=1$. Therefore, the spin connection is
\begin{equation}
\boldsymbol{\omega}_{\;\;\;b}^{a}=\boldsymbol{\alpha}_{\;\;\;b}^{a}+\mathbf{e}^{a}W_{b}-\eta_{bc}\mathbf{e}^{c}W^{a}\label{Spin connection}
\end{equation}

The curvature follows by substituting (\ref{Spin connection}) into
Eq.(\ref{Curvature of Weyl geometry}). After collecting terms and
setting $\mathbf{d}\mathbf{e}^{a}=\mathbf{e}^{b}\land\boldsymbol{\alpha}_{\;\;\;b}^{a}$,
\begin{eqnarray*}
\boldsymbol{\mathcal{R}}_{\;\;\;b}^{a} & = & \mathbf{d}\boldsymbol{\omega}_{\;\;\;b}^{a}-\boldsymbol{\omega}_{\;\;\;b}^{c}\land\boldsymbol{\omega}_{\;\;\;c}^{a}\\
 & = & \left(\frac{1}{2}R_{\;\;\;bde}^{a}\left(\alpha\right)-\left(\delta_{d}^{a}\delta_{b}^{f}-\eta_{bd}\eta^{af}\right)\left(D_{e}W_{f}-W_{e}W_{f}+\frac{1}{2}\eta_{ef}W^{2}\right)\right)\mathbf{e}^{d}\land\mathbf{e}^{e}
\end{eqnarray*}
Antisymmetrizing $de$ to remove the basis and contracting, the Ricci
tensor becomes
\begin{eqnarray*}
\mathcal{R}_{be}=\mathcal{R}_{\;\;\;bae}^{a} & = & R_{be}\left(\alpha\right)-2D_{e}W_{b}-\eta_{be}D^{d}W_{d}+2W_{e}W_{b}-2\eta_{be}W^{2}
\end{eqnarray*}
Because the connection is no longer simply $\boldsymbol{\alpha}_{\;\;\;b}^{a}$,
the Ricci tensor acquires an antisymmetric part,
\[
\mathcal{R}_{\left[be\right]}=-2W_{\left[b;e\right]}
\]
In vacuum this must vanish independently. Even when we consider the
full Einstein equation including sources, the symmetry of the energy
tensor implies
\[
W_{\left[b;e\right]}=W_{\left[b,e\right]}=0
\]
Therefore, $W_{a}=\partial_{a}\phi$ for some $\phi$ and the geometry
is \emph{integrable Weyl}. The vector $W_{a}$ may be removed by a
conformal scaling of the solder form.

\section{Summary and discussion}

We have presented both general coordinate and Poincarè gauge theory
demonstrations that when the Palatini variation includes a possible
abelian symmetry, the result is scale covariant general relativity
in an integrable Weyl geometry.

The difference between the usual claims and these results hinges on
the different form of the covariant derivative for abelian symmetries.
For non-abelian symmetries, the covariant derivative takes the form
\[
D_{\alpha}v^{\beta}=\nabla_{\alpha}v^{\beta}+v^{\mu}\Gamma_{\;\;\;\mu\alpha}^{\beta}
\]
However, fields which transform covariantly under an abelian symmetry,
$\chi\rightarrow e^{k\phi}\chi$ are assigned a weight $k$, $\chi_{\left(k\right)}$
and the covariant derivative reflects this,
\[
D_{\alpha}\chi_{\left(k\right)}=\partial_{\alpha}\chi_{\left(k\right)}-kW_{\alpha}\chi_{\left(k\right)}
\]
or for weighted, vector-valued fields $v_{\left(k\right)}^{\beta}$,
\[
D_{\alpha}v_{\left(k\right)}^{\beta}=\nabla_{\alpha}v_{\left(k\right)}^{\beta}+v_{\left(k\right)}^{\mu}\hat{\Gamma}_{\;\;\;\mu\alpha}^{\beta}-kv_{\left(k\right)}^{\beta}W_{\alpha}
\]
where $\hat{\Gamma}_{\;\;\;\mu\alpha}^{\beta}$ provides covariance
under non-abelian transformations and $kW_{\alpha}$ under abelian
transformations.

These considerations affect the Palatini variation whenever the metric
has nonzero weight $w_{g}$, since the covariant derivatives of the
metric and metric determinant become
\begin{eqnarray}
D_{\mu}g_{\alpha\beta} & = & \partial_{\alpha}g_{\mu\nu}-g_{\beta\nu}\hat{\Gamma}_{\;\;\;\mu\alpha}^{\beta}-g_{\mu\beta}\hat{\Gamma}_{\;\;\;\nu\alpha}^{\beta}-w_{g}W_{\alpha}g_{\mu\nu}\label{Derivative of metric}\\
D_{\mu}g & = & \partial_{\alpha}g-4w_{g}W_{\alpha}g_{\mu\nu}\label{Derivative of g}
\end{eqnarray}
respectively. We have shown that the solution for $\hat{\Gamma}_{\;\;\;\mu\alpha}^{\beta}$
is
\begin{equation}
\hat{\Gamma}_{\;\;\;\mu\alpha}^{\beta}=\frac{1}{2}g^{\beta\nu}\left[\left(g_{\nu\mu,\alpha}+w_{g}g_{\nu\mu}W_{\alpha}\right)+\left(g_{\nu\alpha,\mu}+w_{g}g_{\nu\alpha}W_{\mu}\right)-\left(g_{\mu\alpha,\nu}+w_{g}g_{\mu\alpha}W_{\nu}\right)\right]\label{Weyl covariant derivative}
\end{equation}
The Christoffel term has been augmented in each metric derivative
by a the abelian connection vector $w_{g}W_{\alpha}$ so that the
combination gives the non-abelian connection weight zero and $\hat{\Gamma}_{\;\;\;\mu\alpha}^{\beta}$
is the connection of a \emph{Weyl geometry}.

Expressing the covariant derivative of the determinant of the metric
in terms of the derivative of the metric, we showed that $W_{\alpha}$
must be a gradient, so the Palatini variation yields an \emph{integrable
Weyl geometry}.

We have shown that these same conclusions apply to Poincarè gauge
theory, even though the independent fields become the solder form
and the spin connection, and the Cartan structure equations leave
no room to modify the fields. By systematic determination of the weights
of all relevant fields in terms of the weight $w_{g}$ of the metric,
and careful treatment of the integration by parts, we show that the
torsion acquires a nonvanishing piece
\begin{equation}
\mathbf{T}^{a}=\frac{1}{2}w_{g}\mathbf{e}^{a}\land\boldsymbol{\omega}\label{Torsion giving Weyl}
\end{equation}
This changes the Cartan structure equation to that of a Weyl geometry.

It is interesting to note that when we use $g_{\alpha\beta}$ and
$\hat{\Gamma}_{\;\;\;\beta\nu}^{\alpha}$ as independent variables
the torsion vanishes by the symmetry of the connection and the Weyl
vector emerges as a limited case of non-metricity. By contrast, using
the orthonormal variables $\left(\mathbf{e}^{a},\boldsymbol{\omega}_{\;\;\;b}^{a}\right)$
leads to vanishing non-metricity from the antisymmetry of the spin
connection while the Weyl vector emerges as a limited form of the
torsion. Thus, the Weyl connection is exactly that generalization
of the connection that may be interpreted as \emph{either} non-metricity
or torsion, or equivalently the Weyl connection is the intersection
between non-metricity and torsion.

Notice that the specification of a symmetric connection has a certain
ambiguity, because a Riemannian geometry \emph{with} torsion in the
form of Eq.(\ref{Torsion giving Weyl}) is equivalent to a Weyl geometry
\emph{without} torsion. The final Weyl connection is symmetric and
metric compatible.

The Weyl covariant derivative includes the Weyl vector $W_{\alpha}$
in two distinct ways. First, a weight $k$ vector $v_{\left(k\right)}^{\alpha}$,
the covariant derivative is
\begin{eqnarray*}
D_{\alpha}v_{\left(k\right)}^{\beta} & = & \partial_{\alpha}v_{\left(k\right)}^{\beta}+v_{\left(k\right)}^{\mu}\hat{\Gamma}_{\;\;\;\mu\alpha}^{\beta}+kW_{\alpha}v_{\left(k\right)}^{\beta}
\end{eqnarray*}
Second, $\hat{\Gamma}_{\;\;\;\mu\alpha}^{\beta}$ itself includes
additional terms as in Eq.(\ref{Weyl covariant derivative}) to make
the non-abelian part of the connection invariant under abelian transformations.
The weights of these two occurrences of the Weyl vector are generally
different. The $w_{g}$ in $\hat{\Gamma}_{\;\;\;\mu\alpha}^{\beta}$
is the weight of the metric, while the $w_{k}$ is the weight of the
field $v_{\left(k\right)}^{\beta}$.

A third set of independent variables was studied by Einstein in 1925.
In \cite{Einstein}, the metric and the connection are both generalized
to \emph{asymmetric} fields $\mathfrak{g}^{\mu\nu}$ and $\Gamma_{\;\;\;\mu\nu}^{\alpha}$.
This calculation is reproduced in the Appendix, where it is again
seen to lead to Weyl geometry when $\mathfrak{g}^{\mu\nu}$ is symmetric
or its antisymmetric part may be neglected. It is significant to note
that the fully general ansatz for the independent variables leads
to a vector field in addition to the metric and Christoffel connection.

Finally, we compare our treatment with that of two standard references:
the comprehensive text by Misner, Thorne, and Wheeler, and R. M. Wald's
excellent modern approach. 

The argument by Misner, Thorne, and Wheeler MTW begins with what
the authors stress is a \emph{definition}
\begin{equation}
D_{\mu}\sqrt{-g}\equiv\partial_{\mu}\sqrt{-g}-\sqrt{-g}\Gamma_{\;\;\;\alpha\mu}^{\alpha}\label{MTW derivative of volume}
\end{equation}
However, we show how this determinant may be written in terms of the
covariant derivative of the metric and it follows that
\[
\hat{D}_{\mu}\sqrt{-g}=\partial_{\mu}\sqrt{-g}-\sqrt{-g}\hat{\Gamma}_{\;\;\;\beta\mu}^{\beta}-2w_{g}W_{\mu}\sqrt{-g}
\]
Therefore, the definition of \cite{MTW} requires the vanishing of
abelian part of the connection, the Weyl vector. This allows them
to conclude from their variational equation, equivalent to 
\[
D_{\alpha}\left(g^{\beta\rho}\sqrt{-g}\right)=0
\]
that the covariant derivative of the metric vanishes, $D_{\mu}g_{\alpha\beta}=0$.

Understanding of the difference between our result for the Palatini
variation and the usual conclusion of metric compatibility as due
to abelian symmetry was fostered by studying the proof given in \cite{Wald}.
Here, the generalized derivative is written as the sum of the metric
compatible derivative, $\nabla_{\alpha}$,
\begin{equation}
\nabla_{\alpha}g_{\mu\nu}=\partial_{\alpha}g_{\mu\nu}-g_{\beta\nu}\Gamma_{\;\;\;\mu\alpha}^{\beta}-g_{\mu\beta}\Gamma_{\;\;\;\nu\alpha}^{\beta}=0\label{Metric compatible Riemannian derivative}
\end{equation}
and an additional symmetric tensor $C_{\;\;\;\mu\nu}^{\alpha}$, then
varying the additional tensor. In this case, the derivative of a vector
is written as
\begin{equation}
D_{\alpha}v^{\beta}=\nabla_{\alpha}v^{\beta}+v^{\mu}C_{\;\;\;\mu\alpha}^{\beta}\label{Vector derivative}
\end{equation}
and the metric compatibility condition determines $\Gamma_{\;\;\;\mu\alpha}^{\beta}$
to be the Christoffel connection. Eq.(\ref{Vector derivative}) is
the general form of covariant derivative for a nonabelian group, and
should be modified to
\[
D_{\alpha}v_{\left(k\right)}^{\beta}=\partial_{\alpha}v_{\left(k\right)}^{\beta}+v_{\left(k\right)}^{\mu}\Gamma_{\;\;\;\mu\alpha}^{\beta}-kW_{\alpha}v_{\left(k\right)}^{\beta}
\]
for a weight $k$ vector field. The difference in conclusion follows
by replacing Eq.(\ref{Metric compatible Riemannian derivative}) by
the more general form of $D_{\alpha}$ in Eq.(\ref{Derivative of metric}).
\begin{description}
\item [{\medskip{}
Acknowledgement:}] The author thanks an astute referee for pointing
out the advantages of Wald's approach in handling the integration
by parts.
\end{description}

\section*{Appendix I: Einstein's Original Palatini variation}

In 1925, Einstein proposed a unified theory of Gravitation and Electricity
\cite{Einstein}. This was one of many attempts to geometrically unify
the two known interactions of the era. The resulting model represents
a third set of independent variables beyond $\left(g_{\mu\nu},\Gamma_{\;\;\;\mu\nu}^{\alpha}\right)$
and $\left(\mathbf{e}^{a},\boldsymbol{\omega}_{\;\;\;b}^{a}\right)$
because the metric and connection are both generalized to \emph{asymmetric}
fields $\mathfrak{g}^{\mu\nu}$ and $\Gamma_{\;\;\;\mu\nu}^{\alpha}$.
It was in this paper that Einstein carried out the full Palatini variation,
treating $\mathfrak{g}^{\mu\nu}$ and $\Gamma_{\;\;\;\mu\nu}^{\alpha}$
independently. What interests us here is that he also found a Weyl
geometry. 

The basic premise of this model was the Palatini variation, which,
after introducing the connection $\Gamma_{\;\;\;\alpha\beta}^{\rho}$
and the curvature
\begin{eqnarray*}
R_{\;\;\;\mu\nu\beta}^{\alpha} & = & -\frac{\partial\Gamma_{\;\;\;\mu\nu}^{\alpha}}{\partial x_{\beta}}+\Gamma_{\;\;\;\sigma\nu}^{\alpha}\Gamma_{\;\;\;\mu\beta}^{\sigma}+\frac{\partial\Gamma_{\;\;\;\mu\beta}^{\alpha}}{\partial x_{\nu}}-\Gamma_{\;\;\;\mu\nu}^{\sigma}\Gamma_{\;\;\;\sigma\beta}^{\alpha}
\end{eqnarray*}
built from it, Einstein describes as follows\footnote{Any errors in translation are my own. The notation is preserved from
the original, possibly with different indices where the the photocopy
of the original is too blurry.}:
\begin{quotation}
Independently of this affine connection we introduce a contravariant
tensor density $\mathfrak{g}^{\mu\nu}$ whose symmetry we leave undetermined.
From both we build the scalar
\[
\mathfrak{H=}\mathfrak{g}^{\mu\nu}R_{\mu\nu}
\]
and postulate that simultaneous variation of the integral
\[
\mathfrak{J}=\int\mathfrak{H}dx_{1}dx_{2}dx_{3}dx_{4}
\]
with respect to $\mathfrak{g}^{\mu\nu}$ and $\Gamma_{\;\;\;\mu\nu}^{\alpha}$
as independent variables (and not varied on the boundary) vanishes.
\end{quotation}
Below we reproduce the subsequent calculation of \cite{Einstein}
with a few additional comments. A small amount of newer notation is
introduced for clarity.

The variation of $\mathfrak{J}$ with respect to $\mathfrak{g}^{\mu\nu}$
yields the 16 equations
\[
R_{\mu\nu}=0,
\]
the variation of $\Gamma_{\;\;\;\mu\nu}^{\alpha}$ next gives 64 equations,
\begin{eqnarray}
0 & = & \mathfrak{g}_{\quad,\alpha}^{\mu\nu}+\mathfrak{g}^{\beta\nu}\Gamma_{\;\;\;\beta\alpha}^{\mu}+\mathfrak{g}^{\mu\beta}\Gamma_{\;\;\;\alpha\beta}^{\nu}-\delta_{\alpha}^{\nu}\left(\mathfrak{g}_{\quad,\beta}^{\mu\beta}+\mathfrak{g}^{\rho\beta}\Gamma_{\;\;\;\rho\beta}^{\mu}\right)-\mathfrak{g}^{\mu\nu}\Gamma_{\;\;\;\alpha\beta}^{\beta}\label{Field equation from connection}
\end{eqnarray}

The calculation continues by taking the two traces that do not require
the inverse. These lead to
\begin{eqnarray}
0 & = & \mathfrak{g}_{\quad,\alpha}^{\alpha\nu}-\mathfrak{g}_{\quad,\alpha}^{\nu\alpha}\label{Trace 1}\\
0 & = & -3\left(\mathfrak{g}_{\quad,\alpha}^{\mu\alpha}+\mathfrak{g}^{\alpha\beta}\Gamma_{\;\;\;\alpha\beta}^{\mu}\right)-\mathfrak{g}^{\mu\alpha}\left(\Gamma_{\;\;\;\alpha\beta}^{\beta}-\Gamma_{\;\;\;\beta\alpha}^{\beta}\right)\label{Trace 2}
\end{eqnarray}
Next, the inverse is defined with reversed index order $\mathfrak{g}_{\alpha\beta}$,
so that
\[
\mathfrak{g}^{\alpha\nu}\mathfrak{g}_{\alpha\beta}=\delta_{\beta}^{\nu}=\mathfrak{g}^{\nu\alpha}\mathfrak{g}_{\beta\alpha}
\]
Using this, the third contraction of the field equation Eq.(\ref{Field equation from connection})
with $\mathfrak{g}_{\mu\nu}$ gives 
\begin{eqnarray*}
0 & = & \mathfrak{g}_{\mu\nu}\mathfrak{g}_{\quad,\alpha}^{\mu\nu}+\Gamma_{\;\;\;\beta\alpha}^{\beta}-3\Gamma_{\;\;\;\alpha\beta}^{\beta}-\mathfrak{g}_{\mu\alpha}\left(\mathfrak{g}_{\quad,\beta}^{\mu\beta}+\mathfrak{g}^{\rho\beta}\Gamma_{\;\;\;\rho\beta}^{\mu}\right)
\end{eqnarray*}

The divergence $\mathfrak{g}_{\quad,\beta}^{\mu\beta}$ may be written
in terms of the determinant. We know that
\[
\partial_{\alpha}\sqrt{-\mathfrak{g}}=-\frac{1}{2}\sqrt{-\mathfrak{g}}\mathfrak{g}_{\mu\nu}\partial_{\alpha}\mathfrak{g}^{\mu\nu}
\]
so that
\begin{eqnarray}
\partial_{\alpha}\ln\left(\sqrt{-\mathfrak{g}}\right) & = & \frac{1}{2}\mathfrak{g}^{\mu\nu}\mathfrak{g}_{\mu\nu,\alpha}\label{Determinant relation}
\end{eqnarray}
Also using Eq.(\ref{Trace 2}) Einstein defines a vector density
\begin{equation}
\mathfrak{f}^{\mu}\equiv\frac{1}{3}\mathfrak{g}^{\mu\alpha}\left(\Gamma_{\;\;\;\alpha\beta}^{\beta}-\Gamma_{\;\;\;\beta\alpha}^{\beta}\right)=-\left(\mathfrak{g}_{\quad,\alpha}^{\mu\alpha}+\mathfrak{g}^{\alpha\beta}\Gamma_{\;\;\;\alpha\beta}^{\mu}\right)\label{Vector density}
\end{equation}
Notice that the vector density $\mathfrak{f}^{\mu}$ is proportional
to the trace of the torsion and vanishes for a symmetric Riemannian
connection.

Using the determinant relation Eq.(\ref{Determinant relation}) and
Eq.(\ref{Trace 2}), write the third contraction as
\begin{eqnarray*}
0 & = & \mathfrak{g}_{\mu\nu}\mathfrak{g}_{\quad,\alpha}^{\mu\nu}+\Gamma_{\;\;\;\beta\alpha}^{\beta}-3\Gamma_{\;\;\;\alpha\beta}^{\beta}-\mathfrak{g}_{\mu\alpha}\left(\mathfrak{g}_{\quad,\beta}^{\mu\beta}+\mathfrak{g}^{\rho\beta}\Gamma_{\;\;\;\rho\beta}^{\mu}\right)\\
 & = & -2\left(\partial_{\alpha}\ln\left(\sqrt{-\mathfrak{g}}\right)+\Gamma_{\;\;\;\alpha\beta}^{\beta}\right)+\left(\Gamma_{\;\;\;\beta\alpha}^{\beta}-\Gamma_{\;\;\;\alpha\beta}^{\beta}\right)+\mathfrak{g}_{\mu\alpha}\mathfrak{f}^{\mu}
\end{eqnarray*}
Raising the index and substituting the vector density for the trace
of the torsion, we find a third expression for the vector density.
\begin{equation}
\mathfrak{f}^{\mu}=-\mathfrak{g}^{\mu\alpha}\left(\partial_{\alpha}\ln\left(\sqrt{-\mathfrak{g}}\right)+\Gamma_{\;\;\;\alpha\beta}^{\beta}\right)\label{Vector density 2}
\end{equation}
or, lowering an index, $\mathfrak{g}_{\mu\alpha}\mathfrak{f}^{\mu}=-\left(\partial_{\alpha}\ln\left(\sqrt{-\mathfrak{g}}\right)+\Gamma_{\;\;\;\alpha\beta}^{\beta}\right)$.

Now using \ref{Vector density}, the full equation takes the form
\begin{eqnarray*}
0 & = & \mathfrak{g}_{\quad,\alpha}^{\mu\nu}+\mathfrak{g}^{\beta\nu}\Gamma_{\;\;\;\beta\alpha}^{\mu}+\mathfrak{g}^{\mu\beta}\Gamma_{\;\;\;\alpha\beta}^{\nu}-\mathfrak{g}^{\mu\nu}\Gamma_{\;\;\;\alpha\beta}^{\beta}+\delta_{\alpha}^{\nu}\mathfrak{f}^{\mu}
\end{eqnarray*}
This is Eq.(10) in \cite{Einstein}. We also still have
\begin{eqnarray*}
0 & = & \mathfrak{g}_{\quad,\alpha}^{\alpha\nu}-\mathfrak{g}_{\quad,\alpha}^{\nu\alpha}
\end{eqnarray*}

Finally, we convert the tensor densities to tensors by defining
\[
\mathfrak{g}_{\alpha\beta}=\frac{g_{\alpha\beta}}{\sqrt{-g}}
\]
It follows that $\mathfrak{g}=\frac{1}{g}$ and therefore
\[
\mathfrak{g}_{\alpha\beta}=g_{\alpha\beta}\sqrt{-\mathfrak{g}}
\]

Lower indices of the full equation by contracting with $\mathfrak{g}_{\mu\rho}\mathfrak{g}_{\sigma\nu}$
\begin{eqnarray*}
0 & = & -\mathfrak{g}_{\sigma\rho,\alpha}+\mathfrak{g}_{\mu\rho}\Gamma_{\;\;\;\sigma\alpha}^{\mu}+\mathfrak{g}_{\sigma\nu}\Gamma_{\;\;\;\alpha\rho}^{\nu}-\mathfrak{g}_{\sigma\rho}\Gamma_{\;\;\;\alpha\beta}^{\beta}+\mathfrak{g}_{\sigma\alpha}\mathfrak{g}_{\mu\rho}\mathfrak{f}^{\mu}
\end{eqnarray*}
Now we substitute to eliminate the densities. Using Eq.(\ref{Vector density 2})
and dividing out the determinant yields the final form of the field
equation
\begin{eqnarray}
0 & = & -g_{\sigma\rho,\alpha}+g_{\mu\rho}\Gamma_{\;\;\;\sigma\alpha}^{\mu}+g_{\sigma\nu}\Gamma_{\;\;\;\alpha\rho}^{\nu}+g_{\sigma\rho}\phi_{\alpha}+g_{\sigma\alpha}\phi_{\rho}\label{Reduced equation}
\end{eqnarray}
where we define the vector
\begin{eqnarray*}
\phi_{\rho} & \equiv & -g_{\mu\rho}\sqrt{-\mathfrak{g}}\mathfrak{f}^{\mu}
\end{eqnarray*}
We also still have
\begin{eqnarray*}
0 & = & \mathfrak{g}_{\quad,\alpha}^{\alpha\nu}-\mathfrak{g}_{\quad,\alpha}^{\nu\alpha}
\end{eqnarray*}
Along with 16 equations from the variation of $\mathfrak{g}^{\alpha\nu}$,
and 64 from the connection variation, $\phi_{\alpha}$ must already
be determined. In fact, it is the trace of the torsion.

The remaining parts of \cite{Einstein} consider special cases. First,
Einstein shows that if $\phi_{\alpha}=0$ and $\mathfrak{g}^{\alpha\beta}$
is symmetric we arrive at general relativity. The paper concludes
with a perturbative study focussing on the antisymmetric part of $\mathfrak{g}^{\alpha\beta}$. 

Here we digress, noting that if we take $\mathfrak{g}^{\alpha\beta}$
symmetric but do \emph{not} set $\phi_{\alpha}$ to zero, the connection
acquires an antisymmetric piece. Solving Eq.(\ref{Reduced equation})
with $g_{\alpha\beta}$ symmetric, by cycling indices and combining
in the usual way yields pure-trace torsion
\[
T_{\sigma\alpha\rho}=\Gamma_{\sigma\alpha\rho}-\Gamma_{\sigma\rho\alpha}=g_{\sigma\alpha}\phi_{\rho}-g_{\rho\sigma}\phi_{\alpha}
\]
We have seen that this leads to Weyl geometry, with its symmetric,
metric compatible connection.

\section*{Appendix II: Weyl Weights}

An action functional with an abelian symmetry will have weight zero.
From this and the weight $w_{g}$ of the metric, the weights of all
other fields follow. Only the fields have weight; coordinates do not.
The purely numerical arrays $\eta_{\mu\nu}$ and $\varepsilon_{\mu\nu\alpha\beta}$
have zero weight.

Since we know the relationship between the metric and the components
of the solder form, we have
\begin{eqnarray*}
w\left[g_{\mu\nu}\right] & = & w\left[e_{\mu}^{\;\;\;a}e_{\nu}^{\;\;\;b}\eta_{ab}\right]\\
w_{g} & = & 2w_{e}+w_{\eta}
\end{eqnarray*}
and since $w_{\eta}=0$\footnote{If not, suppose $w\left(\eta_{ab}\right)=w_{\eta}\neq0$. Then we
may always define a weight zero \emph{flat} metric by setting $\tilde{\eta}_{ab}=\left(-\det\eta_{ab}\right)^{-1/4}\eta_{ab}$.}, the weight of the component matrix of the solder form is $w_{e}=\frac{1}{2}w_{g}$.
Since the coordinates have zero weight,
\[
w\left(\mathbf{e}^{a}\right)=w_{e}+w\left(\mathbf{d}x^{\mu}\right)=\frac{1}{2}w_{g}
\]
The volume form is
\begin{eqnarray}
\boldsymbol{\Phi} & = & \,^{*}1\nonumber \\
 & = & \frac{1}{4!}\mathbf{e}^{a}\wedge\mathbf{e}^{b}\wedge\mathbf{e}^{c}\wedge\mathbf{e}^{d}e_{abcd}\nonumber \\
 & = & \frac{1}{4!}\mathbf{d}x^{\mu}\wedge\mathbf{d}x^{\nu}\wedge\mathbf{d}x^{\alpha}\wedge\mathbf{d}x^{\beta}\sqrt{-g}\varepsilon_{\mu\nu\alpha\beta}\label{Weight of volume form-1}
\end{eqnarray}
so its weight is $2w_{g}$. We also need the weight of the orthonormal
Levi-Civita tensor. We have
\[
\mathbf{e}^{a}\wedge\mathbf{e}^{b}\wedge\mathbf{e}^{c}\wedge\mathbf{e}^{d}e_{abcd}=\mathbf{d}x^{\mu}\wedge\mathbf{d}x^{\nu}\wedge\mathbf{d}x^{\alpha}\wedge\mathbf{d}x^{\beta}\sqrt{-g}\varepsilon_{\mu\nu\alpha\beta}
\]
where $\varepsilon_{\mu\nu\alpha\beta}$ is a purely numerical array
and the coordinates are of weight zero. With $w\left(\sqrt{-g}\right)=2w_{g}$
we have
\begin{eqnarray*}
w\left(\mathbf{e}^{a}\wedge\mathbf{e}^{b}\wedge\mathbf{e}^{c}\wedge\mathbf{e}^{d}e_{abcd}\right) & = & 2w_{g}\\
4\cdot\frac{w_{g}}{2}+w\left(e_{abcd}\right) & = & 2w_{g}
\end{eqnarray*}
and therefore $w\left(e_{abcd}\right)=0$. 

To find the weight of the spin connection, consider the structure
equation
\[
\mathbf{d}\mathbf{e}^{a}=\mathbf{e}^{b}\land\boldsymbol{\omega}_{\;\;\;b}^{a}+\mathbf{T}^{a}
\]
The weight of $\mathbf{d}\mathbf{e}^{a}$ is
\begin{eqnarray*}
w\left(\mathbf{d}\mathbf{e}^{a}\right) & = & w\left(\mathbf{d}x^{\mu}\partial_{\mu}\mathbf{e}^{a}\right)\\
 & = & w\left(\mathbf{e}^{a}\right)
\end{eqnarray*}
while
\[
w\left(\mathbf{e}^{b}\land\boldsymbol{\omega}_{\;\;\;b}^{a}\right)=w\left(\mathbf{e}^{b}\right)+w\left(\boldsymbol{\omega}_{\;\;\;b}^{a}\right)
\]
Equating these gives
\begin{eqnarray*}
w\left(\boldsymbol{\omega}_{\;\;\;b}^{a}\right) & = & 0\\
w\left(\mathbf{T}^{a}\right) & = & w\left(\mathbf{e}^{a}\right)=\frac{1}{2}w_{g}
\end{eqnarray*}
For the components of the spin connection,
\begin{eqnarray*}
0 & = & w\left(\boldsymbol{\omega}_{\;\;\;b}^{a}\right)\\
 & = & w\left(\omega_{\;\;\;bc}^{a}\mathbf{e}^{c}\right)\\
 & = & w\left(\omega_{\;\;\;bc}^{a}\right)+w\left(\mathbf{e}^{c}\right)
\end{eqnarray*}
so that $w\left(\omega_{\;\;\;bc}^{a}\right)=-w\left(\mathbf{e}^{c}\right)$.

Then for
\[
\boldsymbol{\Sigma}=\delta\omega_{\quad\mu}^{ab}e_{e}^{\;\;\;\mu}\mathbf{e}^{e}\wedge\mathbf{e}^{c}\wedge\mathbf{e}^{d}e_{abcd}
\]
we wedge with a solder form and extract the volume form $\boldsymbol{\Phi}$,
\begin{eqnarray*}
\mathbf{e}^{f}\wedge\boldsymbol{\Sigma} & = & \delta\omega_{\quad e}^{ab}\mathbf{e}^{f}\wedge\mathbf{e}^{e}\wedge\mathbf{e}^{c}\wedge\mathbf{e}^{d}e_{abcd}\\
 & = & -\delta\omega_{\quad e}^{ab}e^{fecd}e_{abcd}\boldsymbol{\Phi}\\
 & = & 2\delta\omega_{\quad e}^{ab}\left(\delta_{a}^{f}\delta_{b}^{e}-\delta_{a}^{e}\delta_{b}^{f}\right)\boldsymbol{\Phi}\\
 & = & 4\delta\omega_{\quad e}^{fe}\boldsymbol{\Phi}
\end{eqnarray*}
Taking the weight of both sides,
\begin{eqnarray*}
w\left(\mathbf{e}^{f}\wedge\boldsymbol{\Sigma}\right) & = & w\left(\delta\omega_{\quad e}^{fe}\boldsymbol{\Phi}\right)\\
w\left(\mathbf{e}^{f}\right)+w_{\Sigma} & = & w\left(\delta\omega_{\quad e}^{fe}\right)+w\left(\boldsymbol{\Phi}\right)\\
w_{\Sigma} & = & -2w\left(\mathbf{e}^{f}\right)+w\left(\boldsymbol{\Phi}\right)\\
 & = & -w_{g}+2w_{g}
\end{eqnarray*}
and we conclude that $w_{\Sigma}=w_{g}$.

Collecting these results:
\[
\begin{array}{ccccccc}
w\left(\mathbf{e}^{a}\right) & = & \frac{1}{2}w_{g} &  & w\left(\boldsymbol{\omega}_{\;\;\;b}^{a}\right) & = & 0\\
w\left(\boldsymbol{\Phi}\right) & = & 2w_{g} &  & w\left(\omega_{\;\;\;bc}^{a}\right) & = & -\frac{1}{2}w_{g}\\
w\left(e_{abcd}\right) & = & 0 &  & w\left(\mathbf{T}^{a}\right) & = & \frac{1}{2}w_{g}\\
w\left(\eta_{ab}\right) & = & 0 &  & w_{\Sigma} & = & w_{g}
\end{array}
\]
The final equality, $w_{\Sigma}=w_{g}$, is important for carrying
out integration by parts with a covariant derivative.

\end{document}